\documentclass[amsmath,amssymb,nofootinbib,notitlepage,superscriptaddress,twocolumn]{revtex4-2}

\usepackage{esint,graphicx}

\usepackage{amsmath,amsfonts,amsthm,bm}

\usepackage{stackengine}

\usepackage{rotate}
\usepackage{color}

\newcommand{\pd}{\partial}

\DeclareFontFamily{OT1}{pzc}{}
\DeclareFontShape{OT1}{pzc}{m}{it}%
            {<-> s * [1.10] pzcmi7t}{}
\DeclareMathAlphabet{\mathscr}{OT1}{pzc}%
                                {m}{it}

\definecolor{RedWine}{rgb}{0.743,0,0}
\definecolor{RoyalBlue}{rgb}{0.25,.41,.88}
\definecolor{ForestGreen}{rgb}{.13,.54,.13}

\usepackage[backref,breaklinks,colorlinks,citecolor=ForestGreen]{hyperref}

\newcommand{\be}{\begin{equation}}
\newcommand{\ee}{\end{equation}}
\newcommand{\bea}{\begin{eqnarray}}
\newcommand{\eea}{\end{eqnarray}}
\def\ba#1\ea{\begin{align}#1\end{align}}

\newcommand{\scri}{\mathcal{I}}

\begin{document}

\title{Frames and Slicings for  Angular Momentum in Post-Minkowski Scattering}

\author{Samuel E. Gralla}
\author{Kunal Lobo}
\author{Hongji Wei}
\affiliation{Department of Physics, University of Arizona, Tucson, Arizona 85721, USA}

\begin{abstract}
In relativistic physics, angular momentum is paired with a lesser known conserved quantity, the ``mass moment'', which appears as the time-space components of the angular momentum tensor.  Calculations of mass moment in electromagnetic and gravitational scattering of point particles have led to some puzzling behavior in which the radiated mass moment does not appear to match the corresponding mechanical change.  We review the issues and show how the freedoms of time slicing and asymptotic frame may be used to bring all known results into agreement.  The key points are to use hyperboloidal time slices and to allow the perturbative and asymptotic frames to differ by an independent Bondi-Metzner-Sachs (BMS) transformation at early and late times.  The relevant BMS transformation involves a translation found recently by Riva, Vernizzi, and Wong.  Building on this work, we conjecture a flux balance law for all orders in the post-Minkowski expansion.
\end{abstract}

\maketitle

\section{Introduction}

The perturbative description of small-angle gravitational scattering is known as the post-Minkowksian (PM) expansion.  Recently there has been much interest in PM scattering due to connections with fundamental and astrophysical questions---see Ref.~\cite{Damour:2016gwp} and many subsequent references.  As part of this program, some subtleties have arisen regarding the notion of angular momentum \cite{Damour:1981bh,Damour:2019lcq,Damour:2020tta,Gralla:2021qaf,Saketh:2021sri,Gralla:2021eoi,Manohar:2022dea,DiVecchia:2022owy,Veneziano:2022zwh,Bini:2022wrq,DiVecchia:2022piu,Riva:2023xxm}.  Some of the confusion surrounds the time-space components of the angular momentum tensor, or ``mass moment''.  In particular, the change in mechanical mass moment of the particles \cite{Gralla:2021qaf} (or, \textit{scoot}\footnote{The mass moment is equal to the energy times the center of energy evaluated at some canonical time.  A change in mechanical mass moment occurs when a person scoots forward on the floor, moving her center of mass forward.}) has been found to disagree with the mass moment radiated during the scattering encounter \cite{Manohar:2022dea}.  The disagreement occurs at 1PM order, where the particles scoot but there is no balancing radiation at all, and also at 2PM (and presumably higher orders), where the mechanical scoot disagrees with the radiative flux.  However, it is not totally clear that the mechanical and radiative calculations use compatible definitions and gauge conditions, so it is not obvious even whether agreement should be expected.

The electromagnetic (EM) analog problem has provided some intuition \cite{Saketh:2021sri,Gralla:2021eoi}.  The 1PM disagreement may be resolved by realizing that the EM field makes \textit{non-radiative} contributions to the initial and final mass moment \cite{Gralla:2021eoi}, balancing the mechanical scoot.  However, it is difficult to pursue a similar resolution in the gravitational case, since the gravitational field does not have a well-defined stress-energy tensor.  In this paper we will provide an alternative treatment of angular momentum in EM scattering that generalizes better to the gravitational case, and discuss an approach to the gravitational problem that restores agreement at least through 2PM.  

 The main important point is to move away from constant-$t$ slices and instead formulate conservation laws with hyperboloidal slices, as in the definition of timelike infinity in general relativity \cite{1998JMP....39.6573G,1989CQGra...6.1075C,1982RSPSA.381..323P,chakraborty_supertranslations_2022,Compere:2023qoa}.  This means that when calculating initial/final particle contributions, we add together the particle mass moments at equal proper time (instead of equal coordinate time).  In the EM case, we check non-perturbatively that this eliminates non-radiative scoots and restores agreement between the mechanical change and the radiated mass moment, and we demonstrate the agreement explicitly through 2PM.  In the gravitational case, we find that analogous 2PM agreement is obtained between mechanical calculations \cite{Gralla:2021qaf}\footnote{We also discovered a computational error in Ref.~\cite{Gralla:2021qaf}, which has now been corrected in the arXiv version.  Agreement requires both correcting the error and changing to hyperboloidal slicing.} and pseudotensor \cite{Manohar:2022dea} or amplitudes-inspired \cite{DiVecchia:2022owy} definitions of angular momentum flux.
 
The pseudotensor and amplitdes-inspired definitions of angular momentum rely on a choice of a preferred background flat metric.  It is also interesting to consider the Bondi-Metzner-Sachs (BMS) framework \cite{bms1,bms2,1962PhRv..128.2851S}, which only requires asymptotic flatness.  However, in this case there is an ambiguity in relating coordinate choices in the PM calculation (``PM frame'') with coordinate choices in the asymptotic analysis (``BMS frame'').  One can make a natural identification in any region where the particles are widely separated (i.e., at initial or final times), but this identification is \textit{not} preserved by time-evolution.  In other words, if the PM and BMS frames are identified at early times, they can still differ by a BMS transformation at late times.  Ref.~\cite{Riva:2023xxm} has found a formula that we interpret as the translation and supertranslation required for the BMS and PM frames to remain the ``same'' at late times.  The transformation accounts for the  discrepancy between BMS and PM fluxes, bringing all approaches into agreement.

This paper is organized as follows.  In Sec.~\ref{sec:angular-momentum} we review conserved quantities in special relativity and discuss the freedom of integration surface.  For the EM problem, we discuss the early/late field contributions in Sec.~\ref{sec:EM}, the corresponding mechanical contributions in Sec.~\ref{sec:EMmech}, and compute the 2PM scoot (change in mass moment) in Sec.~\ref{sec:2PMEM}. For the gravitational problem, we provide an overview in Sec.~\ref{sec:gravity} and consider the BMS framework in PM scattering in  Sec.~\ref{sec:BMSPM}.  We use units with $c=G=4\pi \epsilon_0=1$.

\section{Conservation laws in special relativity}\label{sec:angular-momentum}

In this section we review angular momentum in special relativity and discuss two natural choices for formulating its global conservation.  Given a conserved stress-energy tensor $T^{\mu\nu}$ (satisfying $\nabla_\mu T^{\mu \nu}=0$) and a Killing field $\xi^\mu$ (satisfying $\nabla_\mu \xi_\nu + \nabla_\nu \xi_\mu=0$), we may construct a conserved current $T^{\mu \nu}\xi_\mu$,  
\begin{align}
\nabla_\nu (T^{\mu \nu}\xi_\mu) = 0.
\end{align}
Integrating this equation over some spacetime four-volume and using Stokes theorem, we find
\begin{align}\label{cons}
\int_{\mathcal{B}} T_{\mu \nu}\xi^\mu n^\nu \sqrt{|h|} d^3x= 0,
\end{align}
where $n^\mu$ is the unit normal to the boundary $\mathcal{B}$ of the volume, which has induced metric $h$ and area element $\sqrt{|h|} d^3x$.\footnote{We assume that all parts of the boundary are either timelike or spacelike; $n^\mu$ is outward with $n^\mu n_\mu=+1$ for timelike portions, and inward with $n^\mu n_\mu=-1$ for spacelike portions.}    We will consider two choices for the boundary $\mathcal{B}$.

\subsection{The box}\label{sec:box}

\begin{figure}
    \centering
    \includegraphics[width=\linewidth]{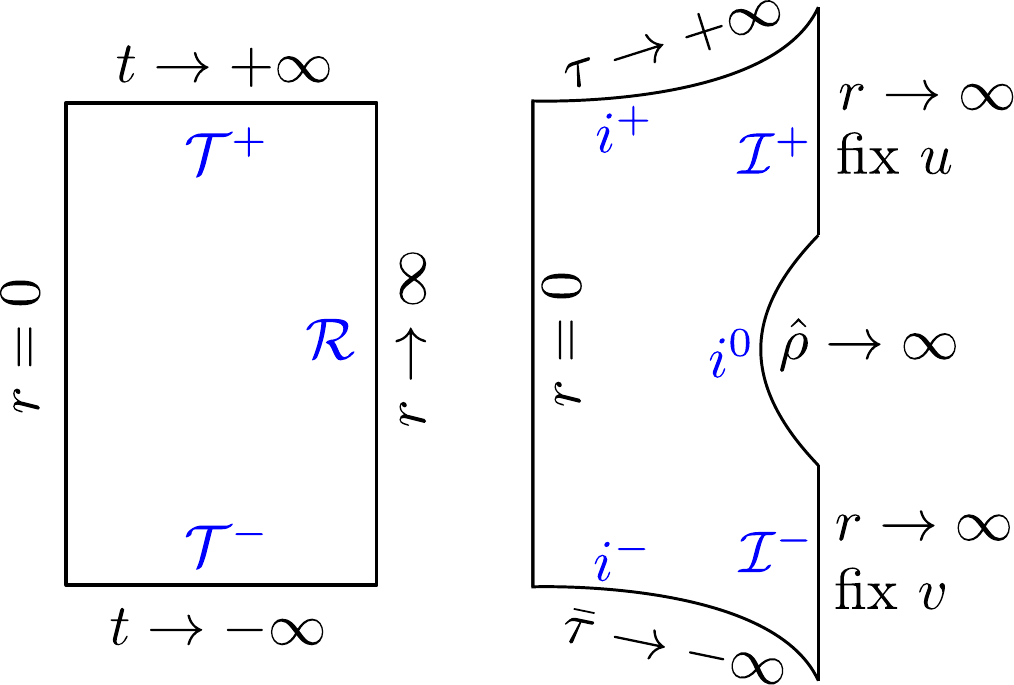}
    \caption{The box (left) and the puzzle piece (right) choices for the integration boundary $\mathcal{B}$ in expressing conservation laws.  The box is drawn taller than it is wide, since one takes $t \to \pm \infty$ at a (slightly) faster rate than $r \to  \infty$. The puzzle piece involves five separate limits, reviewed in the text.}
    \label{fig:box-puzzle}
\end{figure}

The simplest choice for $\mathcal{B}$ consists of initial and final $t={\rm const}$ surfaces bounded by $r= {\rm const}$ spheres at large radius, together with a timelike surface consisting of the sphere cross time.  In a one-dimensional projection this has the appearance of a box (Fig.~\ref{fig:box-puzzle} left).

We will focus on the time-space components of the angular momentum tensor, otherwise known as the ``mass moment''.  These components are accessed by choosing boost Killing fields for $\xi^\mu$,
\begin{align}
    \xi_{(i)} = x^i \frac{\pd}{\pd t} + t \frac{\pd}{\pd x^i},
\end{align}
where $i=1,2,3$.  When evaluated on a spacelike surface $\Sigma$, these give the components of the mass moment vector $N^i$ at the ``time'' represented by $\Sigma$,
\begin{align}\label{N}
    N^i_{(\Sigma)} = \int_\Sigma T_{\mu \nu} \xi^\mu_{(i)} \hat{n}^\nu \sqrt{|h|} d^3x,
\end{align}
where $\hat{n}^\nu$ is the \textit{future-directed} unit normal.  For a $t={\rm const}$ surface this becomes
\begin{align}\label{Nt}
    N^i_{(t)} & = \int(T_{00} x^i + T_{i0} t)d^3x \\
    & = \int \mathcal{E} x^i d^3x - p^i t,
\end{align}
where $\mathcal{E}=T^{00}$ is the energy density and $p^i=\int T^{i0}d^3x$ is the total momentum.  We therefore recognize the mass moment as equal to the value of the total energy times the center of energy at time $t=0$.  

When evaluated on a timelike surface, the integral of \eqref{cons} has the interpretation of (minus) the total flux of mass moment leaving the system, which we denote by $\Delta N^i$.  For an $r={\rm const}$ surface we we have
\begin{align}\label{F}
    \Delta N^i = \int T_{\mu j} \xi^\mu_{(i)} n^j r^2 d\Omega dt.
\end{align}
where $d\Omega=\sin \theta d\theta d\phi$ is the area element on the unit two-sphere and $n^i=x^i/r$ is the outward normal.

To make the box precise we must specify the order of limits for $t \to \infty$ and $r \to \infty$.  Different choices correspond to different accounting systems for the conserved quantities.  The conceptually familiar choice is to capture particles on the bottom and top (representing  initial and final state) and radiation on the sides (representing flux leaving the system).\footnote{Alternatively, one could capture both particles and radiation on the bottom/top, thinking of the radiation as part of the initial/final state.}  This requires a careful simultaneous limit, as now describe.

We describe the top/bottom of the box by $t=\pm t_0$ and the edge by $r=r_0(t_0)$, where $r_0(t_0)$ is some function.  As $t_0 \to \infty$, the box must get wider fast enough that it outruns all massive particles, ensuring that they are captured on the top (and similarly for particles entering from the bottom).  However, it must also get taller fast enough to capture radiation at arbitrarily late times (and analogously for early times).  These requirements may be simultaneously met by taking the ``corner velocity'' $r_0'(t_0)$ to asymptotically approach $1$ from below.  The light moves faster than the corners and ends up on the side; the particles move slower and end on on the top/bottom.  For definiteness we can choose $r_0=t_0-\sqrt{t_0}$.  Thus we define
\begin{itemize}
    \item \textit{Final surface} $\mathcal{T}^+$: The constant-$t$ surface with $r < t_0-\sqrt{t_0}$ as $t_0 \to \infty$.
    \item \textit{Radiation surface} $\mathcal{R}$: The constant-$r$ surface with $r = t_0-\sqrt{t_0}$ as $t_0 \to -\infty$.
    \item \textit{Initial surface} $\mathcal{T}^-$: The constant-$t$ surface with $r < t_0-\sqrt{t_0}$ as $t_0 \to -\infty$.
\end{itemize}

Evaluating Eq.~\eqref{cons} on the box yields three contributions,
\begin{align}\label{cons-t}
    N^i_{\mathcal{T}^+} - N^i_{\mathcal{T}^-}  = \Delta N^i_{\mathcal{R}},
\end{align}
where the subscript $\mathcal{T}_+/\mathcal{T}_-$ refers to Eq.~\eqref{Nt} using the final/initial surface, while the subscript $\mathcal{R}$ refers to Eq.~\eqref{F} on the radiation surface.  We thus have a conservation law stating that the change in mass moment is balanced by radiated mass moment.  This is a formalization of the traditional procedure for understanding the flow of conserved quantities in scattering problems.


\subsection{The puzzle piece}\label{sec:puzzle-piece}

In the relativity literature, asymptotic conservation laws are usually formulated in a different way, involving five asymptotic regions, each with a separate adapted coordinate system.  These may be defined relative to spherical coordinates $(t,r,\theta,\phi)$ as 

\begin{itemize}
\item \textit{Future timelike infinity}, denoted $i^+$, is the limit $\tau \to \infty$ fixing $(\rho,\theta,\phi)$, where \begin{align} t=\tau \cosh \rho, \qquad r=\tau \sinh \rho. \label{i+trans} \end{align}  This limit tracks outgoing particles with proper time $\tau$, constant rapidity $\rho >0$, and constant spatial direction $(\theta,\phi)$.  The points of $i^+$ therefore label outgoing subluminal velocities.  Since $\tau={\rm const}$ surfaces are spacelike hyperboloids, we can think of $i^+$ as a spatial hyperboloid.

\item \textit{Future null infinity}, denoted $\mathcal{I}^+$, is the limit $r \to \infty$ fixing $(u,\theta,\phi)$, where \begin{align} u=t-r.\end{align}  This limit tracks outgoing massless particles with affine parameter $r$, constant retarded time $u$, and constant spatial direction $(\theta,\phi)$.  The points of $\mathcal{I}^+$ therefore label parcels of outgoing radiation. 
  Since $r={\rm const}$ surfaces are timelike cylinders, we can think of $\mathcal{I}^+$ as a timelike cylinder.\footnote{This contrasts with the conformal approach to asymptotics \cite{Penrose1962}, where instead $\mathcal{I}^+$ is a null surface.  However, the two concepts are equivalent---see Ref.~\cite{Compere:2023qoa} for further discussion.}

\item \textit{Spatial infinity}, denoted $i^0$, is the limit $\hat{\rho}\to \infty$ fixing $(\hat{\tau},\theta,\phi)$, where \begin{align}t=\hat{\rho} \sinh{\hat{\tau}}, \qquad  r=\hat{\rho}\cosh{\hat{\tau}}.\end{align}  This limit follows trajectories with velocity larger than light, so that the points of $i^0$ label outgoing superluminal velocities.  Since $\hat{\rho}={\rm const}$ surfaces are timelike hyperboloids, we can think of $i^0$ and a timelike hyperboloid.

\item \textit{Past null infinity}, denoted $\mathcal{I}^-$, is the limit $r \to \infty$ fixing $(v,\theta,\phi)$, where \begin{align} v=t+r. \end{align}  This limit follows  incoming massless particles back in time and is the time-reverse of the limit defining $\mathcal{I}^+$, producing a timelike cylinder whose points label parcels of incoming radiation.

\item \textit{Past timelike infinity}, denoted $i^-$, is the limit $\bar{\tau}\to-\infty$ fixing $(\bar{\rho},\theta,\phi)$, where \begin{align} t=\bar{\tau} \cosh \bar{\rho}, \qquad r=-\bar{\tau} \sinh \bar{\rho}. \label{i-trans}\end{align}  This limit follows incoming massive particles back in time (these have rapidity $\bar{\rho}>0$) and is the time-reverse of the limit defining $i^+$, producing a spatial hyperboloid whose points label incoming subluminal velocities.

\end{itemize}

Drawing together the five surfaces associated with the five limits produces a ``puzzle piece'' shape (Fig.~\ref{fig:box-puzzle} right; see \cite{Compere:2023qoa} for further discussion).  As with the box, we must specify the order of limits.  Consider first the top corner, linking $i^+$ with $\mathcal{I}^+$.  As $\tau_0 \to \infty$ we integrate over some range $\rho<\rho_0(\tau_0)$ for some function $\rho_0(\tau_0)$.  Because $r_0/t_0=\tanh \rho_0$ for the top corner, any function approaching infinity as $\tau \to \infty$ will satisfy the requirement that the corner's velocity asymptotically approaches $1$ from below (see discussion in Sec.~\ref{sec:box} above.)  At future null infinity $\mathcal{I}^+$ one integrates over $u$ at fixed $r$.  The upper cutoff $u_0 = \tau_0 e^{-\rho_0}$ approaches infinity as $\tau_0 \to \infty$ as long as $\rho_0$ approaches infinity slower than logarithmically.  For definiteness, we take $\rho_0(t)=\log \sqrt{\tau_0}$ so that $u_0=\sqrt{\tau_0}$ (much as we took $u_0=\sqrt{t_0}$ in the case of the box).  Similar statements hold for the other corners of the puzzle piece. 

Choosing the puzzle piece for our surface $\mathcal{B}$ in Eq.~\eqref{cons} gives an alternative formulation of conservation laws.  Since $\tau=\sqrt{t^2-r^2}$, the future-directed unit normal to a $\tau={\rm const}$ surface  (with $\tau>0$) is $\hat{n}^\mu=\tau^{-1} (t,x,y,z)$ when expressed in Cartesian coordinates $x^\mu=(t,x,y,z)$.  Using this formula in Eq.~\eqref{N}, the total mass moment on a constant-$\tau$ surface is
\begin{align}\label{Ntau}
N^i_{(\tau)} &= \tau^2 \int d \rho  d\theta d\phi \sinh^2 \!\rho \sin\theta( T_{0 \nu} x^\nu x^i + T_{i \nu} x^\nu t),
\end{align}
which may be compared to Eq.~\eqref{Nt} for a $t={\rm const}$ slice.  Performing the same exercise for the $\bar{\tau}={\rm const}$ surface used at early times, we find the same expression
\begin{align}\label{Ntaubar}
N^i_{(\bar{\tau})} &= \bar{\tau}^2 \int d \bar{\rho}  d\theta d\phi \sinh^2 \!\bar{\rho} \sin\theta( T_{0 \nu} x^\nu x^i + T_{i \nu} x^\nu t).
\end{align}
Now we may define the contributions from the timelike infinities $i^\pm$ to be
\begin{align}
    N^i_{i^+} &= \lim_{\tau \to \infty} N^i_{(\tau)} \label{Ni+}\\ 
    N^i_{i^-} &= \lim_{\bar{\tau} \to -\infty} N^i_{(\bar{\tau})}.\label{Ni-}
\end{align}

The contributions from $\mathcal{I}^\pm$ involve $r={\rm const}$ surfaces and hence take the same form as \eqref{F}, only evaluated in different limits.  In particular, we can define the contributions from the null infinities $\mathcal{I}^\pm$ to be
\begin{align}
    \Delta N^i_{\mathcal{I}^+} & =\lim_{\stackrel{r \to \infty}{{\rm fix } u}}  \int T_{\mu j} \xi^\mu_{(i)} n^j r^2 d\Omega du. \label{Nscriplus} \\
     \Delta N^i_{\mathcal{I}^-} & =\lim_{\stackrel{r \to \infty}{{\rm fix } v}}  \int T_{\mu j} \xi^\mu_{(i)} n^j r^2 d\Omega dv, \label{Nscriminus}
\end{align}
where we remind the reader that $n^i=x^i/r$ is the outward normal to the unit sphere.

The remaining portion of the puzzle piece is spatial infinity $i^0$.  While important in gravitational theory as a mathematical link between $\mathcal{I}^-$ and $\mathcal{I}^+$, it is entirely trivial in electromagnetic scattering.  Neither the particles nor their radiation reach $i^0$, and the Coulomb fields fall off too rapidly to contribute flux (see discussion at the end of Sec.~\ref{sec:EM} below).  The conservation law \eqref{cons} evaluated on the puzzle piece for EM scattering therefore has four contributions,
\begin{align}\label{cons-tau}
    N^i_{i^+} - 
    N^i_{i^-} = \Delta N^i_{\mathcal{I}^-} +  \Delta N^i_{\mathcal{I}^+}.
\end{align}
The left-hand side is interpreted as the change in  mass moment (final $i^+$ minus initial $i^-$), while the right-hand side is the total radiative flux of mass moment, including incoming $\mathcal{I}^-$ and outgoing $\mathcal{I}^+$ radiation.  Eq.~\eqref{cons-tau} for the puzzle piece may be compared with Eq.~\eqref{cons-t} for the box.

\section{EM field contributions}\label{sec:EM}

We now consider the electromagnetic field stress-energy due to point particles moving on some specified worldlines $\bm{r}_I(t)$ with velocity $\bm{v}_I$ and acceleration $\bm{a}_I$.  For each particle we define
\begin{align}
    \bm{R}_I(t) & = \bm{r}-\bm{r}_I(t), \quad \bm{\hat{R}}_I = \bm{R}_I/R_I.
\end{align}
The field is a sum of Coulomb and radiation contributions from each,
\begin{align}\label{FFF}
F^{\mu \nu} = \sum_{a=1}^n \left( F_{(C),a}^{\mu \nu} + F_{(R),a}^{\mu \nu} \right),
\end{align}
whose electric and magnetic fields are
\begin{align}
\bm{E}_{C,I} &= \left. \frac{q_I}{R_I^2} \frac{\bm{\hat{R}}_I-\bm{v}_I}{(1-v_I^2)^{-1}(1-\bm{\hat{R}}_I \cdot \bm{v}_I)^3} \right|_{\rm ret} \label{EC} \\
\bm{E}_{R,I} & = \left.\frac{q_I}{R_I} \frac{ \bm{\hat{R}}_I \times((\bm{\hat{R}}_I-{\bm{v}_I)\times\bm{a}_I)}}{(1-\bm{\hat{R}}_I \cdot \bm{v}_I)^3}\right|_{\rm ret} \label{ER} \\
\bm{B}_I &= \bm{\hat{R}}_I|_{\rm ret} \times \bm{E}_I, \label{BCR}
\end{align}
The subscript ``ret'' indicates evaluation at the retarded time $t_r$ satisfying $R_I(t_r)=t-t_r$.  The Coulomb field falls off like $1/r^2$, while the radiation field falls off like $1/r$.

Now suppose that the particles are widely separated at late times, so that their velocities are asymptotically constant,
\begin{align}\label{vi}
\bm{v}_I & = \bm{V}_I + O(1/t).
\end{align}
Here $\bm{V}_I$ is a constant independent of time.  The position and acceleration then obey
\begin{align}\label{ri}
\bm{r}_I & = \bm{V}_I \ \! t + O(\log t) \\
\bm{a}_I & = O(1/t^2).\label{ai}
\end{align}
The falloff of the acceleration means that the Coulomb field dominates at late times.  The leading contribution is that of a constant-velocity charge.

The integrand \eqref{Nt} for the mass moment on a constant-$t$ surface involves the stress-energy multiplied by distance or time, and integrated over space.  For large $r$ at constant $t$, the stress-energy falls off like $1/r^4$ (from the Coulomb field squared), so the integrand can fall off as slowly as $1/r$.  Thus there can be a contribution from the large-$r$ region as the cutoff $r_0$ is taken to infinity, from a product of two Coulomb fields.  This contribution was computed for two-particle scattering at early and late times in Ref.~\cite{Gralla:2021eoi}.\footnote{Ref.~\cite{Gralla:2021eoi} considered 1PM scattering and computed the cross term (field of particle 1 times field of particle 2) EM contribution to mass moment at early and late times.  As noted in that reference, the effect is not limited to 1PM scattering---it is a general feature of two-particle scattering.  The radiative fields do not contribute on $\mathcal{T}^+$ and $\mathcal{T}^-$ by construction, and for two-particle scattering, we may always boost and rotate so that the particles move in the same direction at asymptotically early or late times. The integrals for the cross-term contribution to the electromagnetic mass moment become identical to those of Appendix.A of Ref.~\cite{Gralla:2021eoi}. 
 The magnitude \eqref{Nscoot} of the mass moment follows from (A15) and (A17) of that reference, noting that $\gamma_1=(m_1+\gamma m_2)/E_0$ and $\gamma_2=(\gamma m_1+m_2)/E_0$ with $E_0^2=m_1^2+m_2^2 + 2\gamma m_1 m_2$.}  At early or late times, in a frame where the particles are asymptotically colinear, the result is 
\begin{align}\label{Nscoot}
    \bm{N}_{\mathcal{T},{\rm EM}\times} = -\frac{q_1 q_2}{\gamma^2 v^2}\log\frac{\gamma_2}{\gamma_1} \hat{\bm{r}}_{12},
\end{align}
where $\hat{\bm{r}}_{12}$ is the unit vector pointing from particle 1 to particle 2.  Here $\mathcal{T}$ represents either $\mathcal{T}^+$ or $\mathcal{T}^-$, $\gamma_I$ are the asymptotic Lorentz factors of the particles $a=1,2$, while $\gamma$ and $v$ are the relative Lorentz factor and velocity.  The notation ``${\rm EM}\times$'' indicates that this is the contribution from the cross-term (particle 1 times particle 2) in the electromagnetic stress-energy.  The formally-infinite self-field terms would require a separate treatment; these contribute to the renormalized center of mass \cite{Gralla:2009md} and are not expected to feature in the electromagnetic sector of the scattering problem.

The purely kinematical contribution \eqref{Nscoot} to the system mass moment must be included at both early and late times in order for mass moment to be globally conserved.  However, it is always canceled by a corresponding mechanical contribution arising from the $\log t$ corrections to the position \eqref{ri} (see Sec.~\ref{sec:EMmech} and Ref.~\cite{Gralla:2021eoi}).  This book-keeping annoyance is unavoidable if $t=\rm{const}$ slices are used.

However, if we instead evaluate the electromagnetic mass moment on a hyperboloidal slice $\tau=\rm{const}$, we find that the late-time integral vanishes---in fact, the \textit{integrand} itself vanishes.  This may be seen directly by constructing the electromagnetic stress-energy from Eqs.~\eqref{FFF}-\eqref{BCR} and using Eq.~\eqref{Ntau} at large $\tau \to \infty$.  We work in Cartesian coordinates but consider $\tau \to \infty$ fixing $(\rho,\theta,\phi)$.  In this case we have $\bm{r}_i \sim \tau$, $\bm{v}_i \sim 1$ and $\bm{a}_i \sim \tau^{-2}$ [similar to Eqs.~\eqref{vi}--\eqref{ai}] as well as $x^i \sim \tau$.  These orderings imply that the Coulomb fields $F^{\mu \nu}_{(C),i}$ scale as $\tau^{-2}$, while the radiation fields scale as $F^{\mu \nu}_{(R),i}$ scale as $\tau^{-3}$.  Given the form of \eqref{Ntau}, only products of Coulomb terms can survive the large-$\tau$ limit.  Plugging in any such product of Coulomb terms, one finds by direct calculation that the integrand of \eqref{Ntau} vanishes.  This calculation can be done most easily by using \eqref{vi} to replace each Coulomb field by the Coulomb field of a constant-velocity charge.

We therefore learn that the electromagnetic field never makes any contribution to the late-time mass moment on the puzzle piece.  An identical argument establishes the same for the early-time mass moment,
\begin{align}
    N_{i^\pm,{\rm EM}} = 0.
\end{align}
Here ``$i^\pm,{\rm EM}$'' refers to the electromagnetic contribution on $i^+$ or $i^-$.\footnote{In the analogous box equation \eqref{Nscoot}, we were forced to restrict to the cross-term contributions, ignoring the infinite self-field contributions.  Such a restriction is not necessary here; the hyperboloidal slices in effect regularize the self-field infinities for mass moment (and also angular momentum).} Working on the puzzle piece, we see that the initial ($i^-$) and final ($i^+$) contributions are purely mechanical.  This conclusion holds for arbitrary $n$-body scattering.

By a similar argument one can check that the ordinary spatial angular momentum integrand also vanishes on $i^\pm$.  The energy and momentum integrands are non-zero, but the (cross-term) integrals vanish.  We therefore conclude that all electromagnetic contributions to conserved quantities vanish on $i^\pm$.

There will, of course, be contributions from the EM stress-energy on $\mathcal{I}^+$: this is the radiative flux that we will consider explicitly in Sec.~\ref{sec:2PMEM} below.  It is easy to check directly from \eqref{EC}--\eqref{BCR} that there are no contributions from $i^0$ or $\mathcal{I}^-$.  The lack of contribution from $\mathcal{I}^-$ is due to the lack of incoming radiation in the retarded fields produced by the particles.

\section{Mechanical Contributions at 1PM}\label{sec:EMmech}

It is instructive to examine the form of the mechanical contributions to mass moment, for both the box and the puzzle piece.  As a definite example we will consider 1PM scattering of point charges with masses $m_1,m_2$ and charges $q_1,q_2$.  From Eqs.~(26)-(29) of Ref.~\cite{Gralla:2021eoi}, the 1PM trajectories (first correction to straight line motion) may be written
\begin{align}
x_1 &=  \frac{m_2\gamma_2}{E_0} b + \frac{q_1 q_2}{b m_1 \gamma_1 v^2}\left( v  t + X_1 \right)  \label{x1} \\
z_1 &=\frac{\gamma  m_2}{\gamma_1 E_0}v \left( t -  \frac{q_1 q_2 E_0}{m_1 m_2 \gamma^3 v^3}\textrm{arctanh} \frac{v t}{X_1} \right) \label{z1} \\
x_2 &=  -\frac{m_1\gamma_1}{E_0} b - \frac{q_1 q_2}{b m_2 \gamma_2 v^2}\left( v  t + X_2 \right)  \label{x2} \\
z_2 &=-\frac{\gamma m_1}{\gamma_2 E_0 }v \left( t -  \frac{q_1 q_2 E_0}{m_1 m_2 \gamma^3 v^3}\textrm{arctanh} \frac{v t}{X_2} \right) \label{z2}
\end{align}
where we define
\begin{align}
    E_0 &= \sqrt{m_1^2 + m_2^2 + 2\gamma m_1 m_2}\label{E0} \\ 
\gamma_1 & = \frac{\gamma m_2 + m_1}{E_0}, \quad
\gamma_2 = \frac{\gamma m_1 + m_2}{E_0} \label{gamma12} \\
X_i & = \sqrt{(\gamma_i^2/\gamma^2) b^2+v^2 t^2} \label{X}.
\end{align}
Here the particles move in the $xz$ plane with initial separation (impact parameter) $b$ in the $x$ direction and initial relative velocity $v$ (relative Lorentz factor $\gamma$) in the $z$ direction.  The initial individual Lorentz factors are denoted $\gamma_1$ and $\gamma_2$, and the initial total energy is denoted $E_0$.  These are also the final Lorentz factors and total energy; there is no radiation or energy exchange at this order.  The small parameters in this expansion are $q_1 q_2/(b m_I)$, where $I=1,2$ labels the particles.  In this section we keep consistently to linear order in these parameters, corresponding to order 1PM in the formal expansion.

We may compute the particle mass moments from these trajectories as 
\begin{align}\label{particle mass moment}
    \bm{N}_I = E_I \bm{r}_I - \bm{p}_I t,
\end{align}
where $E_I$ and $\bm{p}_I$ are the special-relativistic energy and momenta of the point particles $a=1,2$.  We will consider the mass moments of each particle at early and late times.  The transverse ($x$) components reflect the symmetric displacement from the origin in this frame at 1PM order,
\begin{align}
    N_1^x & = -N_2^x  = \frac{\gamma_1 \gamma_2 m_1 m_2}{E_0} , \quad t \to \pm \infty.
\end{align}
The longitudinal ($z$) components are more interesting.  Expanding at early and late times, we have
\begin{align}
N_1^z & \sim \mp \frac{q_1 q_2}{ \gamma^2 v^2}\left(\log  \frac{2 \gamma v |t|}{\gamma_1 b}-1 \right) , \quad t \to \pm \infty \label{N1t} \\
N_2^z & \sim \pm \frac{q_1 q_2}{ \gamma^2 v^2}\left(\log  \frac{2 \gamma v |t|}{\gamma_2 b}-1 \right), \quad t \to \pm \infty.\label{N2t}
\end{align}
Here $\sim$ indicates asymptotic equality, and we have kept all finite and divergent terms.  (The error is $O(|t|^{-1})$.)  Because $\gamma_1 \neq \gamma_2$ in general, the sum $\bm{N}_1+\bm{N}_2$ does not in general vanish at early or late times,
\begin{align}
N^z_1 + N^z_2 = \mp \frac{q_1 q_2}{\gamma^2 v^2} \log \frac{\gamma_2}{\gamma_1}, \qquad t \to \pm \infty 
\end{align}
Noting that $\bm{\hat{r}}_{12} = \mp \bm{\hat{z}}$ as $t\to\pm\infty$, we see that this residual non-zero contribution precisely balances the electromagnetic mass-moment \eqref{Nscoot}, making the total mass moment zero at both early and late times.  However, since the signs of the EM and mechanical contributions both reverse from early to late times, there is net exchange of mass moment between the particles and the field.  Note that this contribution proportional to $q_1 q_2$ is 1PM order: factoring out the dimensions $m_I b$ of mass moment (where $i=1,2$ corresponds to either mass), we see the small parameter $q_1 q_2/(m_I b)$.

Notice that we can make the expressions for $\bm{N}_1$ and $\bm{N}_2$ look more similar by using proper time,
\begin{align}
N_1^z & \sim \mp \frac{q_1 q_2}{ \gamma^2 v^2}\left(\log  \frac{2 \gamma v |\tau_1|}{b}-1 \right), \quad \tau_1 \to \pm \infty \\
N_2^z & \sim \pm \frac{q_1 q_2}{ \gamma^2 v^2}\left(\log  \frac{2 \gamma v |\tau_2|}{b}-1 \right), \quad \tau_2 \to \pm \infty,
\end{align}
where $\tau_1 = t/\gamma_1$ and $\tau_2 = t/\gamma_2$ are proper time parameters for the two particles.  Thus, if we instead add the mass moments together at the same proper time $\tau=\tau_1=\tau_2$, we do find cancellation.  This constant-$\tau$ addition is exactly what we do when evaluating the particles' mechanical stress-energy on the hyperboloid via \eqref{Ntau} and \eqref{i+trans} (or \eqref{Ntaubar} and \eqref{i-trans} at early times).  That is, the hyperboloid provides exactly the needed surface to get the log contributions to mass moment to cancel, and thereby to remove the non-radiative scoot.

\section{2PM EM Scoot}\label{sec:2PMEM}

We now fill a gap in the literature by computing the scoot at 2PM in EM scattering.  We work on the puzzle piece so that the scoot is purely radiative.  We first compute the radiated mass moment at 2PM using the 1PM trajectories presented in the previous section, and then confirm that it matches the change in mechanical mass moment using the 2PM trajectories of Ref.~\cite{Saketh:2021sri}.

According to Eq.~\eqref{Nscriplus}, the flux of mass moment through $\scri^+$ is given integrating $T_{\mu j} \xi^\mu_{(i)} n^j r^2$ over a large sphere at each retarded time $u$.  We again work with Cartesian components but take the limit at fixed $(u,\theta,\phi)$.  The boost Killing field components $\xi^\mu_{(i)}$ grow linearly with $r$ at fixed $u$, so a finite, non-zero contribution in the limit comes from the $1/r^3$ part of the stress-energy tensor.  This means that radiated mass moment comes entirely from cross-terms between $O(1/r^2)$ Coulomb fields and $O(1/r)$ radiation fields, just as occurs with ordinary angular momentum (e.g., \cite{Bonga:2018gzr}).  This contrasts with the radiated energy and momentum, which involve the radiation field squared.  Since the Coulomb field begins at lower PM order than the radiation field, radiation of mass moment and angular momentum occurs at a lower PM order than radiation of energy and momentum \cite{Lapiedra:1979yd,Saketh:2021sri}.

Following standard terminology, the PM order of an expression is equal to the number of powers of the coupling constant ($G$ for gravity, $k=(4\pi \epsilon_0)^{-1}$ for EM) appearing in the expression.  (However, we set these constants to one, so that in practice the PM order is determined by dimensional analysis.)  We will denote terms that scale like $k^n$ and faster as $O(n\text{PM})$.  Since we compute radiation, we will also expand at large $r$ fixing $u$.  In this section, error terms $O(1/r^n)$ refer to $r$ at fixed $u$.

We denote the constant 0PM velocity by $\bm{V}_I$,
\begin{align}
    \bm{r}_I = \bm{V}_I t + O(1\text{PM}).
\end{align}
The 0PM retarded time expanded at large $r$ (fixing $u$) takes a simple form,
\begin{align}\label{trsimp}
    t_r = u \frac{1}{1-\bm{V}_I\cdot \bm{\hat{r}}} + O\left(1\text{PM},\frac{1}{r}\right),
\end{align}
with the notation $O(x,y)=O(x)+O(y)$.
(We drop the label $I$ on the retarded time $t_r$.)
The retarded position is effectively the origin in this limit,
\begin{align}
    \bm{R}_I|_{t_r} = \bm{r} + O(1\text{PM},r^0).
\end{align}
The leading Coulomb field \eqref{EC} is thus
\begin{align}
    \bm{E}_{C,I} &= \frac{q_I(\bm{\hat{r}}-\bm{V}_I)}{r^2 \Gamma_I^2 (1-\bm{V}_I \cdot \bm{\hat{r}})^3} + O\left(2\text{PM}, \frac{1}{r^3}\right) \label{EC1} \\
\bm{B}_{C,I} &= \bm{\hat{r}}  \times \bm{E}_{C,I} + O\left(2\text{PM}, \frac{1}{r^3}\right) ,\label{BC1}
\end{align}
where $\gamma_I=(1-\bm{V}_I)^{-1/2}$ is the (constant) 0PM Lorentz factor already given in Eq.~\eqref{gamma12}. The leading radiation field is given by 
\begin{align}
\bm{E}_{R,I} &= \frac{q_I \bm{\hat{r}} \times((\bm{\hat{r}}-\bm{V}_I)\times\bm{a}_I|_{t_r}) }{ r (1-\bm{V}_I \cdot \bm{\hat{r}} )^3}+ O\left(3\text{PM},\frac{1}{r^2}\right)\label{ER1} \\
\bm{B}_{R,I} &= \bm{\hat{r}} \times \bm{E}_{R,I} + O\left(3\text{PM},\frac{1}{r^2}\right),\label{BR1}
\end{align}
Noting that the flux integral \eqref{F} has a hidden power of $1/k$ in front of it, the 2PM mass moment flux involves only the terms displayed in Eqs.~\eqref{trsimp}--\eqref{BR1}.  It will happen that the $u$-dependence of the integrand comes entirely from the acceleration, so it is helpful to note the $u$-integral of  $\bm{a}(t_r)$.  Using \eqref{trsimp} to change variables, we find
\begin{align}
\int_{-\infty}^{\infty} \bm{a}_I|_{t_r} du & = \Delta \bm{v}_I(1-\bm{V}_I \cdot \bm{\hat{r}}) + O\left(2\text{PM},\frac{1}{r}\right). \label{aa}
\end{align}
where $\Delta \bm{v}_I=\int_{-\infty}^\infty \bm{a}_I(t)dt$ is the change in velocity of particle $I$. Using the trajectories \eqref{x1}--\eqref{z2}, we find
\begin{align}
    \Delta \bm{v}_1 & = \frac{2 q_1 q_2}{b \gamma_1 m_1  v} \bm{\hat{x}} \\
    \Delta \bm{v}_2 & = -\frac{2 q_1 q_2 }{b \gamma_2 m_2  v} \bm{\hat{x}}. \label{Deltav2}
\end{align}

We have now assembled all the ingredients needed to compute the radiative flux \eqref{Nscriplus}.  The Cartesian components of the stress-energy tensor may be constructed using Eqs.\eqref{EC1}--\eqref{BR1} for $\bm{E}$ and $\bm{B}$.  The time integral in \eqref{Nscriplus} can be performed with the aid of \eqref{aa}--\eqref{Deltav2}, and the angular integrals are also straightforward.  This gives the flux of mass moment $\Delta N^i_{\mathcal{I}^+}$.  We also calculate the flux of angular momentum $\Delta L^y_{\mathcal{I}^+}$ [Eq.~\eqref{Nscriplus} using $\xi = -z \pd_x + x \pd_z$] in the same way.  The results are
 \begin{widetext}
 \begin{align} 
\Delta N^x_{\mathcal{I}^+} &= \frac{2 q_1 q_2 }{3 b E_0 m_1 m_2 v}\left(2 \gamma \left(m_2^2 q_1^2-m_1^2 q_2^2\right) +2 m_1 m_2 \left(q_1^2-q_2^2\right)
+\frac{3 q_1 q_2}{\gamma^3 v^3} \left(m_1^2-m_2^2\right)\left( \text{arctanh } v - \gamma^2 v \right) \right) \label{Nfluxx} \\
   \Delta N^z_{\mathcal{I}^+}  &=0 \label{Nfluxz} \\
\Delta L^y_{\mathcal{I}^+} &= \frac{-4 q_1 q_2}{3 b E_0 m_1 m_2 v}\left(\gamma v\left(m_2^2q_1^2+ m_1^2q_2^2\right)+\frac{3 q_1 q_2}{\gamma^2 v^2} m_1 m_2 \left(\text{arctanh } v - \gamma^2 v \right) \right). \label{Lflux}
\end{align}
\end{widetext}
These should each equal the mechanical change in the corresponding conserved quantity, as long as the mechanical contributions are calculated at early/late \textit{proper} times (corresponding to the hyperboloidal slices of the puzzle piece).  The 2PM change in mechanical angular momentum was calculated by Ref.~\cite{Saketh:2021sri} using proper time, and indeed our \eqref{Lflux} agrees with Eq.~(3.16) of that reference.  The change in mechanical mass moment can also be calculated from the trajectories given in Ref.~\cite{Saketh:2021sri}.\footnote{The trajectories given in Eq.~(A25) of Ref.~\cite{Saketh:2021sri} contain typographical errors; the correct expressions were provided to us by M.V.S. Saketh in a private communication.}  We use Eq.~\eqref{particle mass moment} to calculate the individual mass moments of each particle, add them together at the same proper time $\tau$ to find the total mechanical mass moment $N^{\rm mech}(\tau)$, and calculate the difference between the final ($\tau \to \infty$) and initial ($\tau \to -\infty$) values. We find that the change in mass moment matches Eqs.~\eqref{Nfluxx} and \eqref{Nfluxz}.  Thus there is complete agreement between radiated and mechanical contributions through 2PM in EM scattering.

\section{Conservation laws in General Relativity}\label{sec:gravity}

 Conservation laws are more difficult to formulate in general relativity because of the large coordinate freedom and the lack of a local stress-energy tensor for the gravitational field.  To recover familiar special-relativistic notions, one must consider some kind of expansion about flat spacetime.  One can take the expansion to hold globally, as in the PM expansion, or only asymptotically, as needed for non-perturbative scattering.  The latter approach is more general but also comes with a more general freedom of asymptotic frame, including so-called supertranslations in addition to the usual Poincar\'e freedom \cite{bms1,bms2,1962PhRv..128.2851S}.  (The collection of all transformations forms the BMS group.) 
 The supertranslation freedom presents subtleties when comparing the two frameworks, and we will see that even the Poincar\'e degrees of freedom in the frameworks do not necessarily map in the obvious way.  In this section we consider both frameworks in the context of PM scattering and discuss how various choices may be made to bring 2PM results into agreement.

\subsection{PM expansion}\label{sec:PMgrav}

In the PM expansion the bodies are represented as point particles moving on worldlines $z_I^\mu(\tau)$ for $I=1,2$ in a nearly flat spacetime.  The motion of the particles is determined by the geodesic equation (or more general equations, if spin or higher moments are included) in the self-consistent spacetime, with the singular metric perturbations suitably regularized.  We will not review the literature here but merely quote results; we refer the reader to Ref.~\cite{Bel:1981be} for a foundational reference.

The PM expansion has a Poincar\'e freedom associated with the flat background metric as well as the gauge freedom of infinitesimal diffeomorphisms (i.e., diffeomorphisms expanded in $G$ that reduce to the identity as $G\to0$).  The perturbed particle positions are gauge-dependent, but one expects the numerical values to become meaningful in some preferred class of gauges, at least at early and late times when the particles are widely separated.  In particular, one would expect that in such gauges, it is sensible to discuss the initial and final values of the 10 Poincar\'e charges just by using special-relativistic formulae with the particle trajectories $z_I^\mu$.

Using a gauge that arises naturally in self-force calculations (``Lorenz gauge'')\footnote{The Lorenz gauge was used for portions of the calculation involving the linearized Einstein equation; certain non-linear terms were handled using isotropic coordinates for the Schwarzschild metric.}, two of us calculated the full 2PM trajectories and reported on the change in angular momentum and mass moment \cite{Gralla:2021eoi} computed on constant-$t$ slices.  The result contains logarithmic terms precisely analogous to those considered in the EM case above.  However, as in the EM case, these may be eliminated by using constant-$\tau$ slices instead.  Using the trajectories in the appendix of Ref.~\cite{Gralla:2021eoi},\footnote{We have also fixed a computational error in the trajectories of Ref.~\cite{Gralla:2021eoi}, which has the effect of removing the last terms in Eqs.~ (157) and (159) therein.  We have corrected the arXiv version and submitted an erratum.} we find 
\begin{align}
    N^x_{i^+} - N^x_{i^-} & = 2(1+v^2)\frac{\gamma m_1 m_2}{bv^4E_0} (m_1^2-m_2^2) f(v) \label{DeltaNx} \\
    N^z_{i^+} - N^z_{i^-} & = 0 \label{DeltaNz} \\
    L^y_{i^+} - L^y_{i^-}  & = 4(1+v^2)\frac{\gamma^2 m^2_1 m^2_2}{b v^3 E_0} f(v) \label{DeltaLy}
\end{align}
with
\begin{align}
    f(v) \equiv \left(\frac{8}{3}v^3  - v + (1-3v^2)\textrm{arctanh}\ \! v\right).
\end{align}

One would expect that these mechanical changes are balanced by a radiative flux of angular momentum and mass moment, such that the totals are conserved.  However, the definitions of radiated angular momentum and mass moment are more subtle in the gravitational case, since there is no local, gauge-invariant stress-energy tensor for the gravitational field.  We refer the reader to Refs.~\cite{Damour:2019lcq,Mougiakakos:2021ckm,jakobsen:2021smu,Manohar:2022dea,DiVecchia:2022owy,DiVecchia:2022piu,Bini:2022wrq,Riva:2023xxm} for discussions of the various subtleties involved in defining angular momentum in the PM expansion.  Here we will simply note that the 2PM mechanical changes \eqref{DeltaNx}--\eqref{DeltaLy} do match the radiated fluxes computed by these authors; in particular see Refs.~\cite{Manohar:2022dea,DiVecchia:2022owy}.  The use of constant-$\tau$ slices to compute mechanical contributions is essential for this agreement; otherwise one has logarithmic terms in the mechanical results, which do not match any corresponding terms in the fluxes of Refs.~\cite{Manohar:2022dea,DiVecchia:2022owy}. 

Another approach to angular momentum in PM scattering has been taken by Bini and Damour \cite{Bini:2022wrq}.  The idea is to integrate out the electromagnetic field, replacing the local field theory description with a multi-particle action whose interaction terms are non-local in time.  This action then has a full Poincar\'e group of Noether charges, which are conserved for the time-symmetric dynamics.  The charges associated with boosts differ from the mechanical mass moment by interaction terms, which (at early and late times) become precisely logarithmic terms we have emphasized.  The use of this interaction-corrected mass moment is an alternative way to remove the log terms from the mechanical scoot and restore agreement with radiative calculations.

\section{BMS angular momentum in the PM expansion}\label{sec:BMSPM}

The most generally applicable definition of angular momentum in the relativity literature is based on the Bondi framework \cite{bms1,bms2,1962PhRv..128.2851S}.  One introduces an asymptotic structure at null infinity and considers the BMS group of transformations that preserve it.  To each BMS element one may assign a charge $Q(u)$ together with a flux integral encoding its change with time \cite{Dray:1984rfa,Wald:1999wa,Barnich:2011mi}.  The charges associated with BMS elements that are asymptotically rotations may be considered angular momenta.  However, the BMS group is infinite-dimensional (having supertranslations in addition to translations), and there is an infinite number of such ``angular momenta''.  In essence, the freedom of origin in the classical angular momentum is promoted to a whole freedom of a function on the sphere.  Much effort has been devoted to studying angular momentum in this framework; we refer to Ref.~\cite{Flanagan:2015pxa} for a particularly clear presentation.

It is natural to ask how the Bondi framework can account for angular momentum flow in PM scattering, and important steps were taken in Refs.~\cite{Veneziano:2022zwh,Riva:2023xxm}.  In this section we will review these results with a somewhat  different viewpoint, and conclude with a conjecture on how they may generalize to higher PM order.

We use the notation of Ref.~\cite{Compere:2023qoa} except that we reverse the roles of $M$ and $m$: here the Bondi mass aspect is denoted $M$ and the particle masses are denoted $m_n$.  In this coordinate-based approach to the Bondi framework, one has a notion of Bondi coordinates $(u,r,\theta,\phi)$ in which the asymptotic $r\to\infty$ expansion takes a certain form.  This form involves tensors on the two-sphere (indices $A,B,C,\dots)$ that also depend on time $u$.  The physical information is contained in the triple $\{M,C_{AB},N_A\}$, composed of the Bondi mass aspect, shear, and angular momentum aspect, respectively.  To each coordinate transformation preserving the asymptotic form one associates a charge, and after modding out the zero-charge transformations one is left with the BMS group.  For a given spacetime, a definite choice of $\{M,C_{AB},N_A\}$ fixes the BMS freedom; we will say that such a choice constitutes a ``BMS frame''.

The BMS transformations and their charges may be represented by a scalar $T(\theta,\phi)$ and conformal Killing field $Y^A(\theta,\phi)$ on the sphere.  The $\ell=0,1$ harmonics of $T$ correspond to the four spacetime translations and associated energy and momenta, while the higher harmonics correspond to supertranslations and associated supermomenta.  The six different choices of $Y^A$ map to the six rotations/boosts and six different components of the angular momentum tensor.  The map to conserved quantities in our conventions is shown in table \ref{tab:BMS}.  In the simplifying case of a purely electric shear (expected to occur for early and late times), the conserved charges take simple forms (e.g., Eqs.~108-109 of \cite{Compere:2023qoa}),
\begin{align}
    Q_T & = \frac{1}{4\pi}\int_{S^2}M T d\Omega \, , \label{QTscri+} \\
    Q_Y & = \frac{1}{8\pi}\int_{S^2}N_A Y^A d\Omega\,. \label{QYscri+}
\end{align}

We are particularly interested in the dependence of the Lorentz charges $Q_Y$ on the choice of Bondi frame.  Assuming $\dot{C}_{AB}=0$ in addition to purely electric shear (as expected to hold for early and late times), the change in a Lorentz charge $Q_Y$ under a supertranslation $T$ is (see, e.g., Eq.~\eqref{supertranslation NhatA appendix} below) 
\begin{align}\label{deltaTQY}
    \delta_T Q_Y =  \frac{1}{8\pi} \int d\Omega Y^A(3 M \pd_A T + T \pd_A M).
\end{align}
Although this formula is derived for an infinitesimal supertranslation, the supertranslation-invariance of $M$ implies that it holds for a finite supertranslation as well.

\begin{table}
    \centering
    \begin{tabular}{|c|c|c|}
        \hline
        charge & name & generator \\
        \hline
       $E$  &  Energy &$\!\!\!\!\!\!\!\!\!\!\!\!\!\!\!\!\!\!\!\!\!\!\!\!\!\! T = 1$\\
       $\,P^i$ & Momentum &$\!\!\!\!\!\!\!\!\!\!\!\!\!\;  \; T  = n^i(\theta,\phi)$\\
       $\;\,\;P_{\ell m}$ & Supermomentum & $\!\!\!\!\!\!\!\!\!\!\!\!\!\qquad \!\!\!\!\! T = Y_{\ell m}(\theta, \phi)$\\
       $L^i$ & Angular Momentum &$\!\!\!\!\!\!\!\!\!\!\!\!\qquad\;\;\;\; Y^{A} = - \epsilon^{AB} \partial_Bn^i(\theta,\phi) $\\
       $\,N^i$ & Mass Moment &$\qquad \!\!\!\!\!\!\!\!\!\!\!\!\!\!\!\!\!\!\!Y^{A} =  \partial^An^i(\theta,\phi) $\\
       \hline
    \end{tabular}
    \caption{Conventions for BMS Charges and generators. We use the standard orientation  $\epsilon_{\theta\phi}=+\sin\theta$. }
    \label{tab:BMS}
\end{table}

\subsection{Single particle}

Using the Bondi framework to understand PM scattering requires finding an asymptotic coordinate transformation from the PM metric to Bondi coordinates.  This is rather non-trivial at intermediate times, but at early and late times one would expect the metric to behave as some kind of weakly nonlinear superposition of stationary metrics that have been boosted and translated.  To make progress, it is natural to start with Bondi coordinates for a boosted, translated Schwarzschild metric.

There is, of course, a whole BMS freedom in attaching a BMS frame to the Schwarzschild metric.  To reflect the idea of a boosted, translated Schwarzshild metric, we can insist that the 10 Poincar\'e charges match those of a particle in special relativity on a straight-line trajectory,
\begin{align}
    x^i(t) = v^i (t - T) + b^i, \label{pp}
\end{align}  
where $v^i, T$ and $b^i$ are constants.  We use the notation $Q^{\rm mech}$ to denote this collection of charges,
\begin{align}
    &  \quad Q^{\rm mech} = \{E,p^i,L^i,N^i\} \textrm{ with:} \nonumber \\
E & = \gamma m, \qquad \qquad \quad p^i = \gamma m v^i \label{Qmech} \\
L^i & = \gamma m \epsilon^{i}{}_{j k} b^j v^k, \quad N^i = \gamma m (v^i-b^i T). \nonumber
\end{align}
(We only need to match seven charges in order to fix the seven free constants $v^i,b^i,T$.  The other three rotations are absorbed in the vector notation.)

To find a BMS frame for the Schwarzschild metric which has these charges, first note the outgoing Eddington-Finklestein coordinates for a black hole of mass $m$ provides a Bondi frame with
\begin{align}
    M & = m \\
    C_{AB} & = 0 \\
    N_A & = 0.
\end{align}
To make the Poincar\'e charges match, we can perform an asymptotic boost and translation.  We first perform the boost and then the translation.  The relevant formulas were derived in App.~\ref{sec:BMStrans}.  Under the boost, the mass aspect changes to Eq.~\eqref{msolnfinite}, while the shear and angular momentum aspect remain vanishing.  Under the translation, the mass aspect and shear remain invariant, while angular momentum aspect changes according to Eq.~\eqref{supertranslation NhatA appendix}.  Expressing these results covariantly for a boost in the $v^i$ direction, the result is
\begin{align}
    M^{\sf nat} & = \frac{m}{\gamma^3 (1-v_i n^i)^3} \label{Mminimal} \\
    C^{\sf nat}_{AB} & = 0 \label{Cminimal} \\
    N^{\sf nat}_A & = 3M D_A B + B D_A M \label{Nminimal}
\end{align}
with
\begin{align}
    B & = b_i n^i - T.
\end{align}
One can also perform a supertranslation without affecting the Poincar\'e charges, but this just introduces introduces additional complication (non-zero shear) with no clear benefit.  This frame is a ``good cut'' since it has vanishing shear \cite{Newman:1966ub}.

The frame \eqref{Mminimal}-\eqref{Nminimal} is in some sense the most natural Bondi frame for a boosted, translated, spinless particle.  We therefore call it the ``natural'' frame, denoted with superscript ``{\sf nat}''.  The charges of this frame match the mechanical charges,
\begin{align}\label{natismech}
    Q^{\rm mech} = Q^{\sf nat}.
\end{align}
Notice the font distinction: we use a sans-serif font ``{\sf nat}'' to denote a Bondi frame, whereas the original font ``mech'' is used to denote the collection of charges \eqref{Qmech} associated with the point particle \eqref{pp}.  Of course, the Bondi frame {\sf nat} has (infinitely many) more charges than the ten in the collection $Q^{\rm mech}$; here we mean that the Poincar\'e charges match.  The remaining charges (the supermomenta) are just the $\ell \geq 2$ parts of the mass aspect $M^{\sf nat}$ given in \eqref{Mminimal}.

Alternatively, we could begin with a Schwarzschild metric that has been translated and boosted using special-relativistic formulas, and attempt to find a natural asymptotic coordinate transformation to Bondi coordinates.  This was done to $O(1/r)$ in Ref.~\cite{Veneziano:2022zwh}, giving the mass aspect and shear, and to $O(1/r^2)$ in Ref.~\cite{Riva:2023xxm}, providing the angular momentum aspect as well.  The resulting Bondi frame is instead
\begin{align}
    M^{\sf harm} & = \frac{m}{ \gamma^3 (1-v_i n^i)^3} \label{Mharmonic} \\
    C_{AB}^{\sf harm} & = -(2 D_A D_B - \gamma_{AB} D^2)S \label{Charmonic} \\
    N_A^{\sf harm} & = 3M D_A (B+S) + (B+S) D_A M, \label{Nharmonic}
\end{align}
where now we introduce
\begin{align}
    S & = -2 \gamma m(1-v_i n^i) 
    \log\left(\gamma(1-v_i n^i)\right). \label{S} 
\end{align}
Since the calculations of Ref.~\cite{Riva:2023xxm} begin with harmonic gauge, we call this the ``harmonic frame'' (labeled with superscript ``{\sf harm}'').  The harmonic frame satisfies the ``nice cut'' condition \cite{OMMoreschi_1988}: 
\begin{align}
\left( - \frac{1}{4} D^{A} D^{B} C_{AB}^{\sf harm} + M^{\sf harm}_{\ell \geq 2}\right) = 0.
\end{align}

Since the natural and harmonic frames refer to the same (Schwarzschild) spacetime, they must be related by a BMS transformation.  Referring to Eq.~\eqref{supertranslation NhatA appendix}, we see that the difference is a translation and supertranslation encoded in the function $S$. 
 The BMS charges are also in general different, which means that BMS Poincar\'e charges do \textit{not} match the mechanical Poincar\'e charges of the point particle.  Instead, we have
\begin{align}
    Q^{\sf harm} & = Q^{\sf nat} + \delta_S Q = Q^{\rm mech} + \delta_S Q, \label{crazy}
\end{align}
where $\delta_S Q$ is given for Lorentz charges $Q_Y$ in Eq.~\eqref{deltaTQY}.\footnote{Since the mass moment $M$ is invariant under supertranslations, it is the same in the {\sf nat} and {\sf harm} frames.} That is, if Bondi charges are computed in the harmonic frame, then an additional translation and supertranslation are required to to recover the corresponding mechanical charges,
\begin{align}\label{Qmech}
    Q^{\rm mech} = Q^{\sf harm} - \delta_S Q.
\end{align}
This may be compared with Eq.~(3.6) of Ref.~\cite{Riva:2023xxm}, noting that $\delta_T Q_Y$ is equal to $j(M,T)$ in their notation. Ref.~\cite{Riva:2023xxm} proposed this formula as part of a definition of mechanical charges; here we emphasize the interpretation that the harmonic frame is translated and supertranslated relative to the natural frame that inherits the mechanical values.

\subsection{Multiple particles}

We now promote single-particle formulas of the previous section to the multi-particle context in a natural way.  For the mass aspect and shear, which appear at $O(1/r)$ in the Bondi expansion and hence obey linear equations, a simple superposition will suffice.  It is thus natural to expect that ``natural'' Bondi frames ${\sf nat^\pm}$ can be defined at early/late times in $n$-particle scattering by\footnote{Since the particle velocities are finite at late times, the limit in Eq.~\eqref{Mminimal2} is independent of whether we use coordinate time or proper time.  We have expressed the limits as $t \to \pm\infty$ for notational compactness; if we instead used proper time, these would have to be given separately as $\tau \to \infty$ and $\bar{\tau} \to -\infty$.}
\begin{align}
    \lim_{u \to \pm \infty} M^{{\sf nat}{}^\pm} & = \lim_{t \to \pm \infty}\sum_{n=1}^{N}\frac{m_n}{\gamma_n^3 (1-\bm{v}_n \cdot \bm{n})^3}\label{Mminimal2} \\
    \lim_{u \to \pm \infty} C_{AB}^{{\sf nat}{}^\pm} & = 0. \label{Cminimal2}
\end{align}
Here $m_n, \gamma_n, v_n$ refer to the point particle positions in harmonic gauge.   Eqs.~\eqref{Mminimal2} and \eqref{Cminimal2}, promote Eqs.~\eqref{Mminimal} and \eqref{Cminimal}, to the multi-particle context by superposition.

These conditions only partially fix the BMS frame ${\sf nat}^\pm$.  The mass aspect formula \eqref{Mminimal2} fixes the boost and rotation degrees of freedom (given the known values of $m_n$ and $\bm{v}_n$), while the no-shear condition \eqref{Cminimal2} fixes the $\ell \geq 2$ supertranslations. (See Eq.~\eqref{supertranslation C appendix} and \eqref{electric shear appendix} for the transformation law of the shear.)  However, the differential operator in \eqref{Charmonic2} annihilates $\ell=0,1$ modes of $S^{\pm}$, so this condition has no effect on the translation degrees of freedom.  

To fix the translations, one option would be to provide a similar formula for the angular momentum aspect, promoting Eq.~\eqref{Nminimal} to the multi-particle context.  However, since the angular momentum aspect appears at $O(1/r^2)$ where the equations are no longer linear, it is not clear what form a suitable promotion would take. Resolving this question would require finding a coordinate transformation from a multi-particle harmonic-gauge spacetime to a natural Bondi frame, which is a task of considerable complexity.   However, it should still be possible to find a Bondi frame whose charges agree with the mechanical charges, so we can fix the translation degree of freedom of our natural frame by imposing
\begin{align}\label{natismech2}
    Q^{{\rm mech}^\pm} = \lim_{u \to \pm \infty} Q^{{\sf nat}^\pm},
\end{align}
promoting Eq.~\eqref{natismech}.  Here by mechanical charges we mean
\begin{align}
    Q^{{\rm mech}^+} & = \lim_{\tau \to \infty} Q^{\rm mech} \\
    Q^{{\rm mech}-} & = \lim_{\bar{\tau} \to -\infty} Q^{\rm mech},
\end{align}
where we are careful to use the hyperboloidal slicing so of the puzzle piece to define early $(\bar{\tau} \to -\infty)$ and late ($\tau \to \infty$) time limits.  This distinction matters for the Lorentz charges (angular momentum and mass moment).

Eqs.~\eqref{Mminimal2}, \eqref{Cminimal2}, and \eqref{natismech2} define Bondi frames ${\sf nat}^+$ and ${\sf nat}^-$ in terms of the PM trajectories at late and early times, respectively.  Both frames satisfy the good cut condition (vanishing shear).  It is well known that this condition is not preserved under evolution, an effect associated with gravitational memory (e.g., \cite{Strominger:2014pwa}).  Thus these two frames are indeed distinct,
\begin{align}
    ({\sf nat}^+) \neq ({\sf nat}^-).
\end{align}
Instead, the two frames will be related by a BMS transformation.  The gravitational memory induces to a supertranslation, but in principle there could also be translations, rotations, and boosts.

We can analogously define early/late harmonic Bondi frames by promoting Eqs.~\eqref{Mharmonic}, \eqref{Charmonic}, and \eqref{Qmech}, 
\begin{align}
    \lim_{u \to \pm \infty} M^{{\sf harm}{}^\pm} & = \lim_{t \to \pm \infty} \sum_{n=1}^{N}\frac{m_n}{\gamma_n^3 (1-\bm{v}_n \cdot \bm{n})^3} \label{Mharmonic2} \\
    \lim_{u \to \pm \infty} C_{AB}^{{\sf harm}{}^\pm} & = -(2 D_A D_B - \gamma_{AB} D^2)S^\pm, \label{Charmonic2} \\
    \label{Qmech2}
    \lim_{u \to \pm \infty} Q^{{\sf harm}{}^\pm} & = Q^{{\rm mech}{}^\pm} + \lim_{u \to \pm \infty} \delta_{S^{\pm}} Q,
\end{align}
where we now introduce a multi-particle version of $S$, 
\begin{align}
    S^{\pm} = \lim_{t \to \pm \infty} \Big( -2 G\sum_{n=1}^N & m_n \gamma_n (1-\bm{v}_n \cdot \bm{n}) \nonumber \\ & \times \log\left(\gamma_n(1-\bm{v}_n \cdot \bm{n})\right) \Big). \label{S2}
\end{align}
Eqs.~\eqref{Mharmonic2}, \eqref{Charmonic2} and \eqref{Qmech2} define harmonic Bondi frames ${\sf harm}^+$ and ${\sf harm}^-$ in terms of the PM worldlines at late and early times.  The natural and harmonic frames are related at all times $u$ by a translation and supertranslation,
\begin{align}\label{natharm}
    Q_Y^{{\sf nat}^\pm} = Q_Y^{{\sf harm}^\pm} - \delta_{S^\pm} Q.
\end{align}
The change $\delta_S Q(u)$ is given in terms of the mass aspect $M(u,\theta,\phi)$ in Eq.~\eqref{deltaTQY}.  This mass aspect is the same in the harmonic and natural frames (at all times) and is given at early/late times by Eq.~\eqref{Mminimal2} or \eqref{Mharmonic2}.

These definitions allow us to restate a key idea of Ref.~\cite{Riva:2023xxm} as the conjecture that, in contrast to the natural frames, the two harmonic Bondi frames ${\sf harm}{}^+$ and ${\sf harm}{}^-$ are the same at 2PM order,
\begin{align}\label{harmless}
    ({\sf harm}{}^+) = ({\sf harm}{}^-) \qquad \text{at 2PM}.
\end{align}
More explicitly, the conjecture is that Bondi data satisfying the early-time conditions [the minus branch of Eqs.~\eqref{Mharmonic2}, \eqref{Charmonic2} and \eqref{Qmech2}] will evolve to Bondi data satisfying the late-time conditions [the plus branch of the same equations] when the 2PM harmonic-gauge evolution equations are used.  The conjecture has not been fully checked, but it predicts the flux balance laws proposed in Ref.~\cite{Riva:2023xxm}.

We will express the mechanical change in terms of Bondi fluxes in the initial natural frame ${\sf nat^-}$,
\begin{align}\label{fun1}
    \lim_{u \to +\infty} Q^{{\sf nat}^-}_Y - \lim_{u \to -\infty} Q^{{\sf nat}^-}_Y = \mathcal{F}_Y^{{\sf nat}^-},
\end{align}
where the Bondi flux is\footnote{The Bondi flux for Lorentz charges $Q_Y$ is expressed in Eq.~(123) in Ref.~\cite{Compere:2023qoa}. By assuming that we work in vacuum and that there is no magnetic part of shear at early and late times, we use Eqs.~(110), (121), together with the evolution equation for $m$ [Eq.~(111) in Ref.~\cite{Compere:2023qoa}] to arrive at Eq.~\eqref{Fy}.  In this process we drop a term proportional to $-\frac{u}{4} Y^A D_A D_B D_C\dot{C}^{BC}$.  This term vanishes under integration on account of the orthogonality of angular harmonics: $C_{AB}$ is $\ell \geq 2$, while $Y^A$ is $\ell=1$.}
\begin{align}
     \mathcal{F}_Y & =\frac{1}{8\pi} \int_{-\infty}^{\infty} du \int d\Omega Y^A \bigg[ \frac{1}{4} D_B\left(\dot{C}^{BC}C_{CA}\right)\label{Fy}\\ 
     &\qquad +\frac{1}{2}C_{AB}D_C\dot{C}^{BC}+\frac{u}{8} D_A \left(\frac{1}{8}\dot{C}_{BC}\dot{C}^{BC}\right) \bigg].\nonumber
\end{align}

The superscript ${\sf nat}^-$ in \eqref{fun1} indicates to use the shear in the ${\sf nat}^-$ frame. 
This frame is convenient because the Bondi angular momentum flux \textit{vanishes} at 2PM order \cite{Veneziano:2022zwh},
\begin{align}\label{natnoflux}
    \mathcal{F}_Y^{{\sf nat}^-} = O(\text{3PM}).
\end{align}
To see this, first note that there is a hidden prefactor of $1/G$ in \eqref{Fy}, such that the flux is one order lower than the integrand.  The last term contributes to the flux first at 3PM in any gauge since $\dot{C}_{BC}$ is of order 2PM.  The special property of  $\textsf{nat}^-$ is that $C_{AB}$ vanishes in the infinite past $u \to -\infty$, so that $C_{AB}$ itself (and not just $\dot{C}_{AB}$) is of order 2PM.  In this case the first two terms also contribute to the flux starting at 3PM.

From Eq.~\eqref{fun1} using  Eqs.~\eqref{natismech2} and \eqref{natharm}, we find
\begin{align}
    \lim_{u \to +\infty} \left( Q^{{\sf harm}^-}_Y - \delta_{S^-}Q_Y \right) - Q^{{\rm mech}^-}_Y = \mathcal{F}_Y^{{\sf nat}^-}.
\end{align}
The 2PM equivalence of ${\sf harm}^+$ and ${\sf harm}^-$, together with the vanishing of the ${\sf nat}^-$ flux at this order \eqref{natnoflux}, now implies that
\begin{align}
    \lim_{u \to \infty} \left( Q^{{\sf harm}^+}_Y - \delta_{S^-}Q_Y \right) - Q^{{\rm mech}^-}_Y = O(\textrm{3PM}),
\end{align}
and we can use \eqref{natharm} again to write
\begin{align}
    \lim_{u \to \infty} \left( Q^{{\sf nat}^+}_Y + \delta_{S^+}Q_Y - \delta_{S^-}Q_Y \right) - Q^{{\rm mech}^-}_Y = O(\textrm{3PM}).
\end{align}
A second application of Eq.~\eqref{natismech2} then implies
\begin{align}
    Q^{{\rm mech}^+}_Y + \lim_{u \to \infty} \left( \delta_{S^+}Q_Y - \delta_{S^-}Q_Y \right) - Q^{{\rm mech}^-}_Y = O(\textrm{3PM}).
\end{align}
Or, noting the linearity of $\delta_S Q_Y$ in $S$, we have
\begin{align}\label{ireallydontknow}
    Q^{{\rm mech}^+}_Y - Q^{{\rm mech}^-}_Y = - \lim_{u \to \infty} \delta_{\Delta S} Q_Y + O({\rm 3PM}),
\end{align}
where $\Delta S \equiv S^+-S^-$ so that
\begin{align}
    \lim_{u \to \infty}  \delta_{\Delta S} Q_Y =     \frac{1}{8\pi} \int & d\Omega Y^A\big[3 M^+ \pd_A(S^+-S^-) \nonumber \\ & + (S^+-S^-) \pd_A M^+\big],\label{stupidshift}
\end{align}
where $S^\pm$ are defined in \eqref{S2}, and for convenience we denote the late-time mass aspect by $M^+$,
\begin{align}\label{lateM}
    M^+ = \lim_{\tau \to  \infty}\sum_{n=1}^{N}\frac{m_n}{\gamma_n^3 (1-\bm{v}_n \cdot \bm{n})^3}.
\end{align}

The integral in Eq.~\eqref{stupidshift} can be evaluated for each choice of $Y^A$, corresponding to the components of  angular momentum and mass moment.  This calculation was done in Ref.~\cite{Riva:2023xxm} in a different notation [See Eq.~(3.25) therein], but we reproduce it here for clarity and completeness.  The integrand is built from $M^+$, $S^+$, and $S^-$, which depend on the harmonic-gauge trajectories via the initial and final velocities of the particles.  Since $S^\pm$ has a hidden factor of $G$ in front, we will need only the 1PM trajectories to determine the 2PM $S^\pm$.  In the canonical scattering setup we consider (center of energy frame with initial motion along $z$ and transverse separation along $x$), the particle four-momenta $p^\mu_{a,\pm}$ at late ($+$) and early ($-$) times are given by Eqs.~(144) and (145) of Ref.~\cite{Gralla:2021qaf} as
\begin{align}\label{pp}
    p^\mu_{i,+} &= p^\mu_{i,-} + \Delta p_i^\mu,
\end{align}
with
\begin{align}
    p^\mu_{1,-} &= \left( \frac{m_1^2 + \gamma m_1 m_2}{E_0},0,0,\frac{\gamma v m_1 m_2}{E_0} \right)\\
    p^\mu_{2,-} &= \left( \frac{m_2^2 + \gamma m_1 m_2}{E_0},0,0,-\frac{\gamma v m_1 m_2}{E_0} \right) \\
     \Delta p^\mu_1 & = - \Delta p^\mu_2  \\ & = \left(0, - \frac{2m_1m_2\gamma}{bv}(1+v^2),0,0\right)+ O(\text{2PM}).\label{pp1}
\end{align}
Plugging Eqs.~\eqref{pp}--\eqref{pp1} into Eqs.~\eqref{lateM} and \eqref{S2}, we obtain
\begin{align}
    M^+ & = \sum_{i=1,2}\frac{m^4}{p^0_{i,+}-p^z_{i,+}\cos \theta } + O(\text{2PM}) \label{M+} \\
    S^+ - S^- &= - \frac{4m_1 m_2\gamma(1+v^2)}{bv}\sin\theta \cos\phi \nonumber \\&\times \log \left(\frac{m_1 + m_2\gamma(1-v\cos\theta)}{m_2 + m_1\gamma(1-v\cos\theta)}\right) + O(\text{3PM}).\label{S+S-}
\end{align}
Note that $p_{i,+}^\mu=p_{i,-}^\mu$ at the 1PM order displayed in \eqref{M+}; we write $p_{i,+}^\mu$ for aesthetic reasons.  

According to the conjecture \eqref{harmless}, the integral \eqref{stupidshift} should match the mechanical change (with a minus sign--see Eq.~\eqref{ireallydontknow}).  As far as we are aware, the mechanical change in angular momentum has not been computed in harmonic gauge, but results in a related gauge\footnote{Both the harmonic gauge and the Lorenz gauge of \cite{Gralla:2021qaf} involve hyperbolic wave equations, where causality is manifest.} \cite{Gralla:2021qaf} were presented as Eqs.~\eqref{DeltaNx}--\eqref{DeltaLy} above.  Plugging Eqs.~\eqref{M+} and \eqref{S+S-} into Eq.~\eqref{stupidshift} and evaluating the integral for the relevant choices of $Y^A$ (see table~\ref{tab:BMS}), we find an exact match.  This provides support for the conjecture \eqref{harmless}.

\subsection{Discussion}

In this section we have reproduced and repackaged the asymptotic calculations of Ref.~\cite{Riva:2023xxm} and compared with the mechanical calculations of Ref.~\cite{Gralla:2021qaf}.  The main difference of interpretation is that  
Ref.~\cite{Riva:2023xxm} regard the term $\delta_{\Delta S} Q_Y$ as due to a ``static'' contribution to angular momentum, where as we view it the effect of a translation and supertranslation needed to reconcile PM and Bondi results. (The (super)translation is applied at late times $u \to \infty$, but its form is determined non-locally, using early-time information as well.) This interpretation has content: the Bondi and PM definitions of angular momentum are sufficiently distinct that, in principle, there could have been no reconciliation between them, and one would have been forced to ``choose'' between two compelling definitions of angular momentum.  Instead, we find that the two different results are related by a BMS transformation, indicating that the difference is just a subtlety about frames in which the conserved quantities are defined. 

It is natural to consider the physical interpretation of the (super)translation $\Delta S=S^+-S^+$ that relates the natural-frame BMS results and the Lorenz-gauge PM results.  The $\ell \geq 2$ modes (i.e., the pure supertranslations) are precisely those that eliminate the change in Bondi shear due to the linear memory effect, which is the total memory effect at this PM order.  In this context one sometimes says that the BMS frame is \textit{supertranslated} as a result of the scattering process \cite{Strominger:2014pwa}.  We may describe the $\ell=0,1$ modes of $\Delta S$ as  an additional \textit{translation} of the frame as well, but there is a key difference to bear in mind: whereas the supertranslation can be defined intrinsically on $\mathcal{I}^+$, the translation is only  \textit{relative} to the harmonic-gauge PM frame.

In particular, the ``supertranslation of the BMS frame'' due to the passage of radiation can be defined as the supertranslation needed to eliminate the change in Bondi shear.  Physically, nearby freely falling observers can measure that their final relative distances differ from those recorded prior to the passage of the radiation---the gravitational memory.  By contrast, we are unaware of any corresponding intrinsic definition of the ``translation of the BMS frame'' from the Bondi data $\{M,C_{AB},N_A\}$.  Physically, we expect that no such definition will exist, since there is no way for freely falling observes to locally measure whether they have ``moved'' as a result of the passage of the radiation (absolute distances cannot be measured).  Such a translation must be relative to something in the bulk; at present, the best we can say is that the BMS frame is translated relative to the PM frame.

Ascribing physical meaning to the translation requires ascribing physical meaning to the BMS and PM frames.  For the natural BMS frame, we can note that the angular coordinates $(\theta,\phi)$ label freely falling observers in the asymptotically flat region $r \to \infty$ with proper time $u$, and the good cut condition corresponds to initially synchronized clocks.  For the PM frame, we can note that the harmonic (or Lorenz) gauge gives rise to causal propagation through the bulk.   Since both frames have physically appealing properties, the relative translation between the frames may very well have physical significance.  It would be interesting to understand this better.

\subsection{Conjecture}

A related question is whether the conjecture \eqref{harmless} can be promoted to higher PM order.  Could it be that the ${\sf harm^-}$ frame evolves to the ${\sf harm^+}$ at all orders in the PM expansion?  The answer is no: The non-linear memory effect \cite{Christodoulou:1991cr} guarantees that there will be a change in Bondi shear in any process involving radiation of energy.  The ${\sf harm^+}$ and ${\sf harm^-}$ frames must therefore differ at least by the supertranslation generated by (see, e.g., Eq.~(117) of Ref.~\cite{Compere:2023qoa})\footnote{In Eq.~\eqref{SNLM}, it is understood that the inverse operator $[D^2(D^2+2)]^{-1}$ acts on the space of functions with no $\ell=0,1$ modes, and that these modes are removed from $\dot{C}_{AB} \dot{C}^{AB}$.  Thus $S_{\rm NLM}$ contains only $\ell \geq 2$ modes.}
\begin{align}
    S_{\rm NLM} = \frac{1}{2} [D^2(D^2+2)]^{-1} \int \dot{C}_{AB} \dot{C}^{AB} du,\label{SNLM}
\end{align}
with ``NLM'' standing for ``non-linear memory''.  
 However, one may conjecture that this is the only difference,
\begin{align}
    {\sf harm^+} =  \mathcal{S}_{\rm NLM}({\sf harm^-}),
\end{align}
where $\mathcal{S}_{\rm NLM}$ represents the action of the supertranslation generated by \eqref{SNLM}.

This conjecture predicts flux-balance laws for the Lorentz charges.  Repeating the steps leading to \eqref{ireallydontknow}, we see that the updated law is
\begin{align}\label{whyamievendoingthis}
    Q^{{\rm mech}^+}_Y - Q^{{\rm mech}^-}_Y = \mathcal{F}_Y^{{\sf nat}^-} - \lim_{u \to \infty} \delta_{\chi} Q_Y,
\end{align}
where now
\begin{align}
    \chi = S^+ - S^- + S_{\rm NLM}.
\end{align}
In effect, the change in mechanical charges is given by three terms: the flux of angular momentum in the natural frame (the initial good cut whose charges agree with the initial Poincar\'e charges), the correction due to gravitational memory (the $\ell \geq 2$ parts of $S^+-S^-$ together with $S_{\rm NLM}$), and the further correction found by Ref.~\cite{Riva:2023xxm} (the $\ell=0,1$ parts of $S^+-S^-$).   It would be interesting to check this conjecture at higher PM order.

 \appendix

\section*{Acknowledgements}
It is a pleasure to thank Geoffrey Comp\`ere, \'Eanna Flanagan, Massimiliano Riva, and Filippo Vernizzi for helpful conversations.  This work was supported by NSF grants PHY-1752809 and PHY-2309191 to the University of Arizona.

\section{BMS transformation laws}\label{sec:BMStrans}
 In this appendix, we derive simplified formulas for the change in the Bondi data $\{m,C_{AB},N_A\}$ under BMS transformations in some special cases of relevance for deriving Eqs.~\eqref{Mminimal}--\eqref{Nminimal} of the main text.  We rely heavily on the formulas in Ref.~\cite{Flanagan:2015pxa}.  However, note that the angular momentum $N_A$ defined in Ref.~\cite{Flanagan:2015pxa} differs from our $N_A$ by a term $u D_A m$.  We denote the $N_A$ of \cite{Flanagan:2015pxa} by $\tilde{N}_A$, and the relationship to our $N_A$ is
 \begin{align}\label{NNtilde}
     N_A = \tilde{N}_A - u D_A m.
 \end{align}

We adopt the convention where $x^A = \{\theta,\phi\}$ are spherical coordinates, with the two-sphere metric $\gamma_{AB}$. We write the covariant derivative compatible with $\gamma_{AB}$ as $D_A$. Bondi coordinates $\{u,r,x^A\}$, along with the definition of metric components (Bondi data $\{m, C_{AB}, N_A\}$) and conserved charges are along with \cite{Flanagan:2015pxa}.  Symmetrization and antisymmetrization are denoted as $()$ and $[]$ over indices respectively\footnote{Denoting  $P(A_1...A_n)$ as permutations, symmetrized or antisymmetrized tensors are given by
\begin{align}
    T_{(A_1...A_n)} &= \frac{1}{n!}\sum_{P(A_1...A_n)}T_{P(A_1...A_n)}\nonumber\\
    T_{[A_1...A_n]} &= \frac{1}{n!}\sum_{P(A_1...A_n)}(-1)^PT_{P(A_1...A_n)}.\nonumber
\end{align}}
We use dots to indicate a $u$-derivative (such as the Bondi news is denoted as $\dot{C}_{AB}$). We consider the vacuum case $T_{\mu \nu}=0$. 

Any symmetric rank-2 tensor $C_{AB}$ on the sphere can be decomposed into an ``electric part'' $C$ and ``magnetic part'' $\Psi$:
\begin{align}\label{decomp CAB appendix}
    C_{AB} & = (-2D_AD_B + \gamma_{AB} D^2)C + \epsilon_{C(A}D_{B)}D^C \Psi,
\end{align}
where $\epsilon_{AB}$ is the Levi-Civita tensor on the sphere with the convention $\epsilon_{\theta \phi} = \sin\theta$. (See Eq.~(105) in Ref.~\cite{Compere:2023qoa} or Eq.~(2.24) in Ref.~\cite{Flanagan:2015pxa}.) Note that the $\ell = 0, 1$ parts of $C$ and $\Psi$ will not contribute to the expression Eq.~\eqref{decomp CAB appendix}.

Using the commutators of the spherical derivatives together with Riemann curvature of the unit sphere $R_{ABCD} = \gamma_{AC}\gamma_{BD} - \gamma_{AD}\gamma_{BC}$, we can select out the electric and magnetic parts of $C_{AB}$ by applying derivative operators
\begin{align}
    D^AD^B C_{AB} &= - D^2(D^2+2)C\label{electric part selector appendix}\\
    D_{[B}D^CC_{A]C} & = \frac{1}{2}\epsilon_{C[A}D_{B]}D^C(D^2+2)\Psi.\label{magnetic part selector appendix}
\end{align}
We will only consider cases where the magnetic part vanishes and the shear is constant $\dot{C}_{AB}=0$,
\begin{align}\label{electric shear appendix}
    C_{AB} = (-2D_AD_B + \gamma_{AB} D^2)C,
\end{align}
where $C$ is constant in time, $\dot{C} = 0$.   note that $\ell = 0,1$ parts of $C$ do not contribute to $C_{AB}$.

From the vanishing of the Bondi news $\dot{C}_{AB}=0$ and the stress energy $T_{\mu \nu}=0$, it follows from the evolution equation Eq.~(2.11a) in \cite{Flanagan:2015pxa} that $\dot{m} = 0$.  For the remainder of the appendix we thus have
\begin{align}\label{mdotCdot0}
    \dot{m} = \dot{C}_{AB} = 0.
\end{align}
However, the angular momentum aspect $N_A$ will in general depend on time $u$.

\subsection{Finite boost from a good cut}

We now consider the special case where the shear vanishes, known as a ``good cut'',
\begin{align}\label{appendix good cut}
    C_{AB}  = 0.
\end{align}
Noting also that $\dot{m}=0$ \eqref{mdotCdot0}, we may obtain the action of an infinitesimal Lorentz transformation from Eqs.~(2.18a)-(2.18c) in \cite{Flanagan:2015pxa},
\begin{align}
    \delta_Y m &=  \frac{3}{2} m \psi+ Y^A D_A m\label{Lorentz trans m appendix}\\
    \delta_Y C_{AB} & = 0 \label{Lorentz trans CAB appendix}\\
    \delta_Y \tilde{N}_A & = (1+\frac{1}{2}u)\psi \tilde{N}_A + \mathcal{L}_Y\tilde{N}_A + \frac{3}{2}u m D_A \psi, \label{Lorentz trans NA appendix}
\end{align}
where $\mathcal{L}_Y$ is the Lie derivative on the sphere with respect to $Y^A$, and $\psi = D_A Y^A$.  The notation $\delta_Y X$ indicates the change in $X$ after an infinitesimal coordinate transformation generated by the associated vector field $\xi^\mu$ given as Eq.~(2.16) of Ref.~\cite{Flanagan:2015pxa}.  We restricted to a Lorentz transformation by setting $\alpha=0$ in that equation.  We also used the equations of motion for $\tilde{N}_A$ [Eq.~(2.11b) in Ref.~\cite{Flanagan:2015pxa}].

We can now express the change in $N_A$ using Eqs.~\eqref{Lorentz trans m appendix}, \eqref{Lorentz trans CAB appendix}, and \eqref{NNtilde}:
\begin{align}\label{Lorentz trans NhatA appendix}
    \delta_Y N_A = \delta_Y \tilde{N}_A - u D_A \delta m = (\mathcal{L}_Y + \psi) N_A.
\end{align}

We now consider the action of a finite boost acting on an ``initial'' configuration
\begin{align}
    m=M, \quad C_{AB}=0, \quad N_A = 0,
\end{align}
where $M$ is a constant independent of $(\theta,\phi)$ (as well as $u$).  This requires integrating up the equations after choosing $Y^A$ to effect a boost in the $z$ direction,
\begin{align}\label{YAboost}
    Y^A = D^A \cos \theta.
\end{align}
First consider the mass aspect \eqref{Lorentz trans CAB appendix}.  If $w$ parameterizes the finite boost, then $m$ becomes a function $m(w;\theta,\phi)$ satisfying
\begin{align}
    \delta_Y m = - \frac{\pd m}{\pd w}
\end{align}
with the sign chosen so that positive $w$ corresponds to a boost in the $+z$ direction.  The normalization of $Y^A$ in \eqref{YAboost} ensures that $w$ is equal to the rapidity of the finite boost.  We thereby obtain a differential equation for $m$,
\begin{align}
    - \frac{\pd m}{\pd w} = -3 m x + (1-x^2) \frac{\pd m}{\pd x},
\end{align}
where $x=\cos\theta$.  One can check by direct calculation that a solution is
\begin{align}\label{msolnfinite}
    m = \frac{M}{(\cosh w - x \sinh w)^3},
\end{align}
as first shown in Ref.~\cite{bms1}. 
This is the unique solution by the Cauchy–Kovalevskaya theorem.  The covariant version of this expression is Eq.~\eqref{Mminimal}.

Eq.~\eqref{Lorentz trans CAB appendix} shows that the shear is invariant, while Eq.~\eqref{Lorentz trans NhatA appendix} shows that the change in mass aspect $N_A$ is a linear operator acting on $N_A$.  In both cases the unique solution with zero initial data is zero, so the boosted configuration also has $C_{AB}=N_A=0$.

\subsection{(Super)translation}

We now consider a supertranslation acting on a Bondi frame satisfying  \eqref{electric shear appendix} and \eqref{mdotCdot0} (but not assuming the good cut condition).  Choosing $Y = 0$ and $\alpha = T$ in Eq.~(2.13) of Ref.~\cite{Flanagan:2015pxa}, the change of Bondi data can be computed from (2.18a-2.18c) in Ref.~\cite{Flanagan:2015pxa}:
\begin{align}
    \delta_{T} m &= 0\label{supertranslation m appendix}\\
    \delta_{T} C & = T\label{supertranslation C appendix}\\
    \delta_{T} \tilde{N}_A & = 3 m D_A T + T D_A m\label{supertranslation N_A appendix},
\end{align}
where we used the equations of motion for $\tilde{N}_A$ [Eq.~(2.11b) in Ref.~\cite{Flanagan:2015pxa}] and also employed \eqref{magnetic part selector appendix} with $\Psi=0$.  In terms of $N_A$ \eqref{NNtilde} we have
\begin{align}\label{supertranslation NhatA appendix}
    \delta_{T} N_A &= \delta_{T} \tilde{N}_A - u D_A \delta_{T} m  \nonumber\\
    &=  3m D_A T + T D_A m,
\end{align}
i.e., $\delta_T N_A = \delta_T \tilde{N}_A$ since $\delta_T m=0$.  

These infinitesimal results integrate up trivially to finite supertranslations.  First, Eq.~\eqref{supertranslation m appendix} shows that $m$ is invariant under the supertranslation.  Next, Eq.~\eqref{supertranslation C appendix} shows that the infinitesimal change in $C$ does not depend on $C$, so the equation is trivially solved as $C=w T$, if $w$ is the group parameter and $T$ is the representative of the infinitesimal supertranslation.  In this case we simply send $w T \to T$ and express the finite supertranslation by the new $T$.  Similar comments apply to \eqref{supertranslation NhatA appendix}: since $m$ is invariant (independent of $w$), the right-hand-side is similarly independent of $\gamma$.  Thus the formulas for finite supertranslations are identical to the infinitesimal ones.

\bibliographystyle{utphys}
\bibliography{PMscattering.bib}

\end{document}